\documentclass[smallabstract,smallcaptions]{dccpaper}

\usepackage{amsmath}
\usepackage{amssymb}
\usepackage{booktabs}

\usepackage{xcolor}
\definecolor{citecolor}{HTML}{0071bc}

\usepackage[pagebackref,breaklinks=true,colorlinks,citecolor=citecolor,bookmarks=false]{hyperref}

\DeclareMathOperator*{\argmax}{arg\,max}
\DeclareMathOperator*{\argmin}{arg\,min}

\begin{document}

\title
{\large
\textbf{Optimal Transcoding Resolution Prediction for Efficient Per-Title Bitrate Ladder Estimation}
}

\author {
    Jinhai Yang, 
    Mengxi Guo, 
    Shijie Zhao$^{\ast}$, \thanks{$^{\ast}$Corresponding author.}
    Junlin Li, 
    Li Zhang\\
    {\small\begin{minipage}{\linewidth}
    \begin{center}
    \begin{tabular}{ccc}
        Bytedance Inc., Shenzhen, China\\
        Bytedance Inc., San Diego, CA, 92122 USA\\
        \texttt{\{yangjinhai.01,guomengxi.qoelab,zhaoshijie.0526,lijunlin.li,lizhang.idm\}}\\
        \texttt{@bytedance.com}
    \end{tabular}
    \end{center}
    \end{minipage}}
}

\maketitle

\begin{abstract}
Adaptive video streaming requires efficient bitrate ladder construction to meet heterogeneous network conditions and end-user demands.
Per-title optimized encoding typically traverses numerous encoding parameters to search the Pareto-optimal operating points for each video.
Recently, researchers have attempted to predict the content-optimized bitrate ladder for pre-encoding overhead reduction.
However, existing methods commonly estimate the encoding parameters on the Pareto front and still require subsequent pre-encodings.
In this paper, we propose to directly predict the optimal transcoding resolution at each preset bitrate for efficient bitrate ladder construction.
We adopt a Temporal Attentive Gated Recurrent Network to capture spatial-temporal features and predict transcoding resolutions as a multi-task classification problem.
We demonstrate that content-optimized bitrate ladders can thus be efficiently determined without any pre-encoding.
Our method well approximates the ground-truth bitrate-resolution pairs with a slight Bjøntegaard Delta rate loss of 1.21\% and significantly outperforms the state-of-the-art fixed ladder.
\end{abstract}

\section{Introduction}

Recent years have witnessed the flourishing of online video consumption, with network traffic growing faster than connection bandwidth~\cite{cisco2020cisco}.
Facing the challenge of limited bandwidth and network heterogeneity, video service providers have invested significantly in adaptive video streaming for on-demand video services.
Previous works have shown that downscaling prior to encoding and upscaling after decoding can improve the overall rate-distortion (RD) performance compared with transcoding at the original scale~\cite{bruckstein2003down}, especially at lower bitrates~\cite{lin2006adaptive}.
Therefore, MPEG introduced the DASH~\cite{sodagar2011mpeg} standard to encode videos with different encoding profiles configured by a bitrate ladder.
The bitrate ladder is usually defined by a set of bitrate-resolution pairs, which specify the transcoding resolution for each preset bitrate.
Instead of using a fixed bitrate ladder for content-agnostic transcoding, an advanced solution is to construct a per-title optimized content-aware bitrate ladder~\cite{per-title}.
Typically, numerous encoding parameters are used on each video to generate sufficient pre-encoded copies to exhaustively search the convex hull of Pareto-optimal operating points across all RD curves.
This brute-force searching process would be time-consuming and resource-intensive due to unnecessary pre-encodings and quality evaluations.

\begin{figure}[t]
\begin{center}
\begin{tabular}{cc}
\epsfig{height=0.88in,file=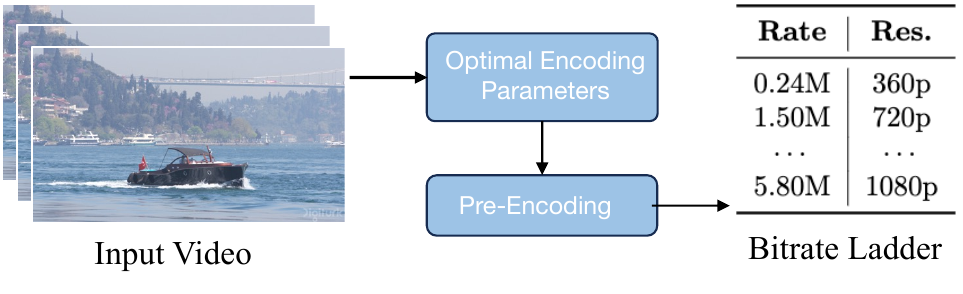} &
\epsfig{height=0.88in,file=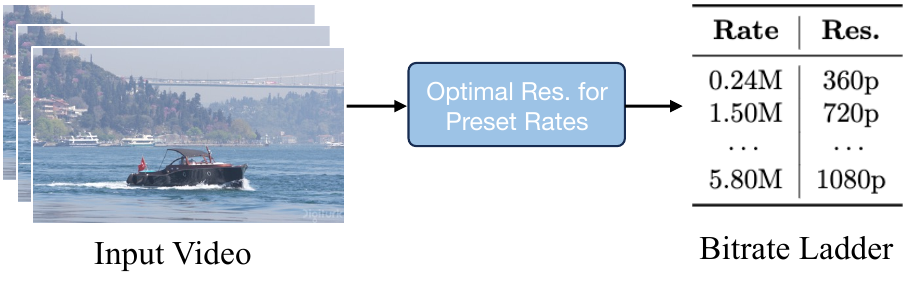} \\
{\small (a) Two-stage bitrate ladder estimation.} & {\small (b) One-stage bitrate ladder estimation.}
\end{tabular}
\end{center}
\vspace{-15pt}
\caption{\label{fig:motivation}%
Comparison between the proposed one-stage bitrate ladder estimation and existing two-stage approaches. 
Our one-stage method eliminates the overhead of pre-encoding.}
\vspace{-5pt}
\end{figure}

To reduce the pre-encoding overhead, recent research has shown interest in predicting the content-aware bitrate ladder using data-driven algorithms~\cite{katsenou2021efficient,paul2022efficient,adhuran2023content}.
However, as shown in Fig.~\ref{fig:motivation}(a), existing works commonly adopt the two-stage paradigm, where the optimal encoding parameter combinations are estimated first and then a reduced set of pre-encodings are conducted to determine the best streaming ladder.
For example, \cite{paul2022efficient} represents possible encoding configurations as a binary matrix whose components express whether the corresponding pairs of resolution and quantization parameter (QP) lie on the Pareto front.
A neural network is employed to predict these optimal combinations from video contents as a multi-label binary classification task to prune unnecessary pre-encodings.
Therefore, such two-stage methods are inefficient and still heavily rely on massive pre-encodings.

In this paper, we propose a novel one-stage framework for efficient bitrate ladder construction.
To determine the appropriate bitrate-resolution pairs, we select a fixed set of bitrates following the multi-codec DASH encoding recipe~\cite{zabrovskiy2018multi} and thereby directly predict the optimal streaming resolution at the predetermined bitrates.
The candidate resolutions are also defined as a finite set of typical transcoding resolutions.
In this way, bitrate ladder estimation can be formulated as a multi-task (in regard to bitrates), multi-class (in regard to resolutions) classification task.
Moreover, this paradigm is also more practical for rate-controlled encoding in bandwidth-limited applications compared with existing QP-controlled solutions~\cite{katsenou2021efficient,katsenou2021vmaf,paul2022efficient}.
Inspired by recent advances in sequence modeling~\cite{chung2014empirical,carreira2017quo,vaswani2017attention}, we develop a Temporal Attentive Gated Recurrent Network (TAGRN) to jointly capture the spatial-temporal features from video clips for optimal transcoding resolution determination.

Experimental results have demonstrated that per-title optimized bitrate ladders can be efficiently constructed without pre-encoding using the proposed method.
For training and evaluation, we encode the large-scale Inter-4K~\cite{stergiou2022adapool} videos with predefined bitrate and resolution ranges.
The quality of encoded videos is evaluated with the perceptual-oriented Video Multi-method Assessment Fusion (VMAF)~\cite{li2016toward} metric to decide the optimal resolutions.
The predicted results closely approximate the ground-truth convex hulls with small Bjøntegaard Delta rate (BD-rate)~\cite{bjontegaard2001calculation} loss and consistently outperform the state-of-the-art fixed ladder~\cite{zabrovskiy2018multi}.

\section{Related Work}
Apart from brute-force searching, there are three common approaches for bitrate ladder construction: model-based, proxy-based, and feature-based.
The model-based methods~\cite{dong2013adaptive,wang2014adaptive} find the optimal sampling ratio by decomposing the overall distortion into the downscaling distortion and the coding distortion via explicit modeling. 
However, these methods tend to be inaccurate in practice due to flawed assumptions about the complex coding systems.
Differently, the proxy-based method~\cite{wu2020fast} performs surrogate pre-encoding with faster encoders to identify the optimal QP-resolution pairs for the target slower encoders.

Our work is most related to the feature-based methods~\cite{katsenou2021efficient,katsenou2021vmaf,paul2022efficient,adhuran2023content}.
In this direction, the spatial-temporal characteristics of video clips have been represented with handcrafted~\cite{katsenou2021efficient,katsenou2021vmaf,adhuran2023content} or learned~\cite{paul2022efficient} features.
Based on handcrafted features, the QP values at the intersection points~\cite{katsenou2021efficient} or knee points~\cite{katsenou2021vmaf} between RD curves of different resolutions were predicted via regression.
Besides, \cite{adhuran2023content} has attempted to estimate the target bitrates corresponding to a set of given quality values.
Differently, a recurrent convolutional network is utilized in \cite{paul2022efficient} to ascertain whether the predefined QP-resolution combinations are Pareto-optimal in an end-to-end manner.
After the determination of the optimal encoding parameters, these approaches still require massive pre-encoding to derive the final bitrate ladder.

In this study, we explore a novel framework to eliminate the pre-encoding burden by estimating the optimal streaming resolutions for preset bitrates in one stage, as shown in Fig.~\ref{fig:motivation}(b).
Our TAGRN extracts spatial features with a frozen 2D convolutional neural network (CNN)~\cite{he2016deep}, which are temporally aggregated via multi-head attention~\cite{vaswani2017attention} and hereafter classified using a gated recurrent unit (GRU)~\cite{chung2014empirical}.

\section{Method}
In this section, we first introduce the preliminary of the proposed one-stage bitrate ladder estimation in Sec.~\ref{sec:formulation}.
Next, we describe the procedure of video encoding and ground-truth construction in Sec.~\ref{sec:groundtruth}.
The details about the structure and training objective of TAGRN are elaborated in Sec.~\ref{sec:network} and Sec.~\ref{sec:objective}, respectively.

\subsection{Problem Formulation}\label{sec:formulation}
The RD characteristic of a sequence in regard to a specified video codec can be derived using constant bitrate (CBR) encoding.
Let $\mathcal{R}$ denote the predefined resolution set whose cardinality $\left|\mathcal{R}\right|=R$, the RD curve of each resolution $r\in\mathcal{R}$ is defined by the bitrate vector $\mathbf{B}=\left[b_1^r,b_2^r,\dots,b_B^r\right]^\top$ and the corresponding quality vector $\mathbf{Q}=\left[q_1^r,q_2^r,\dots,q_B^r\right]^\top$.
To allow for efficient bitrate ladder estimation, the target bitrates are also configured in advance as a fixed set $\mathcal{B}=\{\hat{b}_{i}\}_{i=1}^{B}$.
Using CBR encoding, it is guaranteed that $b_i^r\approx \hat{b}_i$.
The ideal resolution for transcoding at bitrate $\hat{b}_i$ can be denoted by $r_i^*$, where
\begin{equation}\label{eq:labeling}
    r_i^*=\argmax_r q_i^r~.
\end{equation}
In practice, the resulting bitrate $b_i$ after CBR encoding will not be exactly equal to the target $\hat{b}_i$.
Nevertheless, such slight bitrate deviations typically lead to minor differences in quality values and can thus be reasonably ignored if the deviations are small enough.
Therefore, in Sec.~\ref{sec:groundtruth}, we propose a two-step encoding strategy to constrain the bitrate limits for different resolutions to avoid large deviations between the target and the actual bitrates.

\begin{figure}
    \centering
    \includegraphics[width=\linewidth]{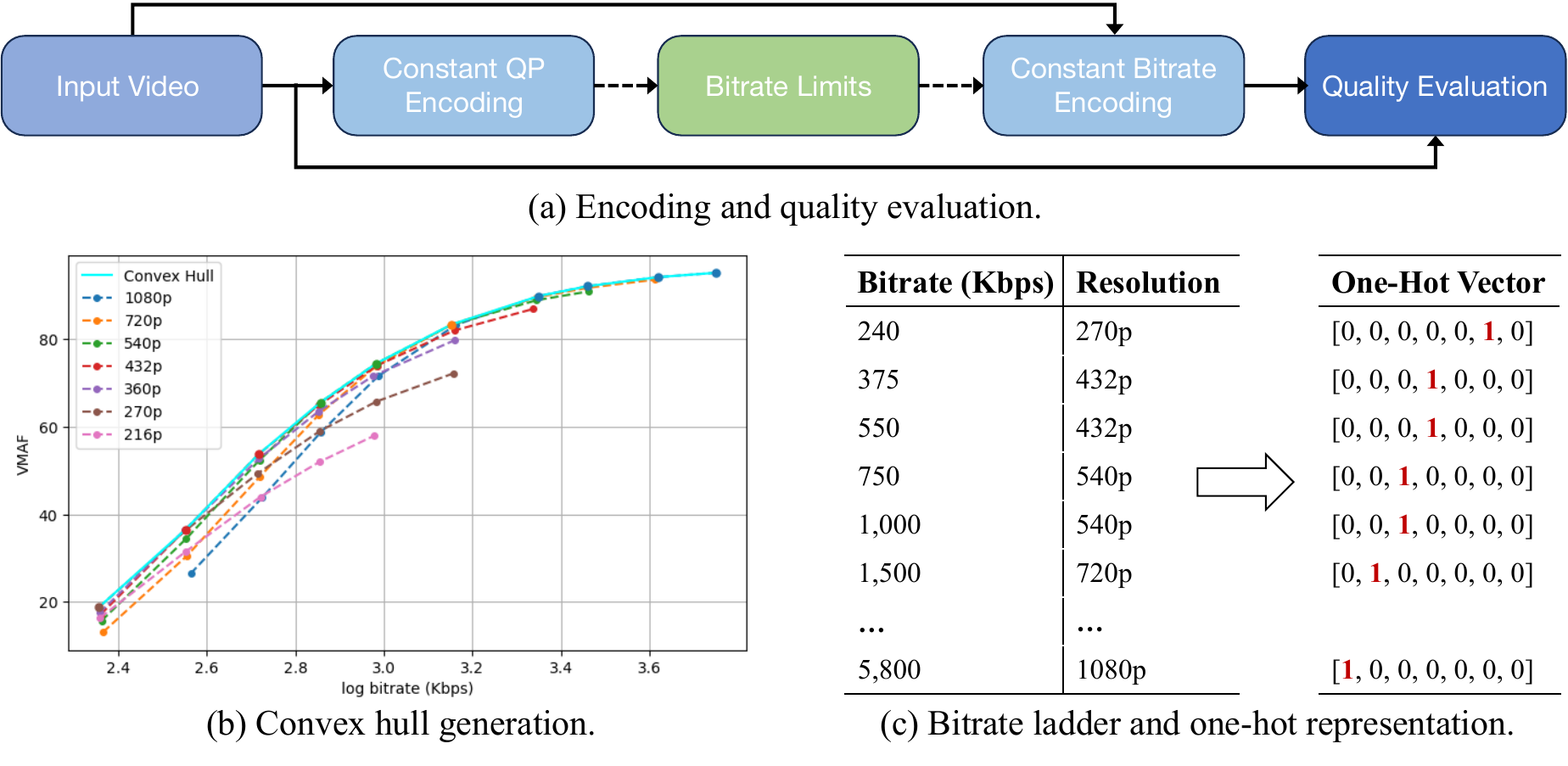}
    \caption{Example of ground-truth bitrate ladder construction and one-hot representation. }
    \label{fig:datalabeling}
\end{figure}

\subsection{Ground-Truth Construction}\label{sec:groundtruth}
We employ a two-step encoding strategy for ground-truth bitrate ladder construction as shown in Fig.~\ref{fig:datalabeling}(a).
Instead of encoding the source videos using the whole set of predefined bitrates for all resolutions, we perform constant QP (CQP) encodings before the CBR encoding step to determine the bitrate limits for each resolution.
More concretely, the videos are encoded at QP=16 and QP=48 to obtain the upper bounds and lower bounds of the target bitrates.
This CQP process provides an appropriate range of bitrates according to the video complexity and specified resolution, which helps minimize the differences between target bitrates and the resulting bitrates.

In this work, all of the video encodings are performed with the x265 encoder.
The two-pass encoding is adopted for stricter rate control.
The encoding resolution covers the typical streaming range from 216p to 1080p following \cite{paul2022efficient}, where $\mathcal{R}=\{1920\times1080,1280\times720,960\times540,768\times432,640\times360,480\times270,384\times216\}$.
The target bitrate set in Kbps is $\mathcal{B}=\{240,375,550,750,1000,1500,2300,3000,4300,5800\}$, following the multi-codec DASH fixed ladder~\cite{zabrovskiy2018multi}.
All the encoded videos are upscaled to the original resolution for quality evaluation.
Notably, all video coding and quality analysis are required only for model training and evaluation but are not necessary at deployment.

An example of convex hull generation and bitrate-ladder representation is shown in Fig.~\ref{fig:datalabeling}(b)(c).
Due to the two-step encoding strategy, the low-resolution versions (\textit{e.g.} 216p) of this sequence are not encoded at high bitrates (\textit{e.g.} $>1000$ Kbps).
The optimal streaming resolutions $\{r_i^*\}_{i=1}^{B}$ are determined with Eq.~\ref{eq:labeling} and then represented with one-hot vectors $\mathbf{Y}=\left[\mathbf{y}_1^*,\mathbf{y}_2^*,\dots,\mathbf{y}_B^*\right]^\top\in\mathbb{R}^{B\times R}$ as the ground-truths.

\subsection{Temporal Attentive Gated Recurrent Network}\label{sec:network}

\begin{figure}
    \centering
    \includegraphics[width=\linewidth]{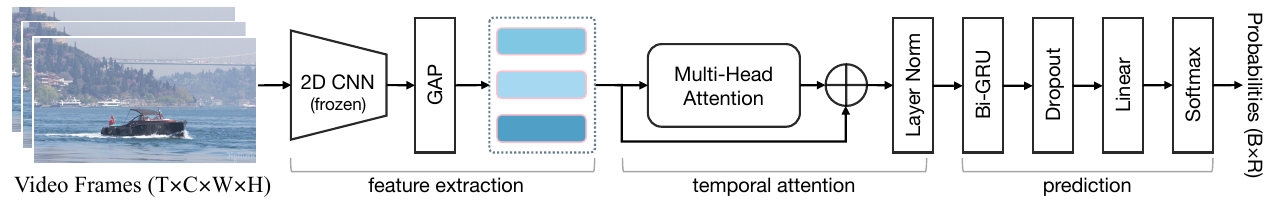}
    \caption{Illustration of the proposed TAGRN. The frame-level spatial features are extracted independently and then cross-frame fused via a temporal attention module for prediction.}
    \label{fig:method}
\end{figure}

The sheer volume of video data makes it a formidable challenge to design an efficient model with respect to understanding the spatiotemporal scene complexity for optimal resolution classification.
Towards addressing this challenge, we develop a temporal attentive gated recurrent network (TAGRN) as illustrated in Fig.~\ref{fig:method}.
In previous work~\cite{paul2022efficient}, only the luma channels of the first 3 frames with a 5-frame temporal stride for each video are processed since a sophisticated network is trained from scratch, and the source videos with shot changes are manually removed.
Differently, given a video with frame size $W\times H$, we uniformly sample $T$ frames to form a chunk that covers most of the relevant information over the entire duration including the scene changes.
The frames are processed in the RGB space (\textit{i.e.} $C=3$) to fully leverage the prior knowledge of 2D CNN obtained by pre-training on large-scale image datasets.
The output feature maps are passed through a global average pooling (GAP) layer to obtain the feature sequence $\mathbf{X}\in\mathbb{R}^{T\times D}$.
The parameters of the 2D CNN can be frozen during training to alleviate the memory and computation requirements.

The temporal relations are also critical to analyzing videos, especially for a codec-related problem.
Therefore, we aggregate the static features in the time domain with the multi-head attention mechanism~\cite{vaswani2017attention}.
The feature sequence is linearly mapped to the query embedding $\mathbf{Q}\in\mathbb{R}^{T\times D}$, key embedding $\mathbf{K}\in\mathbb{R}^{T\times D}$ and value embedding $\mathbf{V}\in\mathbb{R}^{T\times D}$, respectively.
In multi-head attention, these embeddings are divided into $n$ groups with feature dimensions thereby reduced and concatenated after individual self-attentions, followed by a linear projection.
Let $\mathbf{Q}_i\in\mathbb{R}^{T\times \frac{D}{n}}$, $\mathbf{K}_i\in\mathbb{R}^{T\times \frac{D}{n}}$, $\mathbf{V}_i\in\mathbb{R}^{T\times \frac{D}{n}}$ denote the $i$-th group of embeddings, the self-attention output $\mathbf{A}_i\in\mathbb{R}^{T\times \frac{D}{n}}$ is:
\begin{equation}
\mathbf{A}_i=\operatorname{softmax}(\mathbf{Q}_i \mathbf{K}_i^{\top}/{\sqrt{D}}) \mathbf{V}_i~.
\end{equation}

Following the multi-head attention, the temporally attended feature $\mathbf{A}\in\mathbb{R}^{T\times D}$ is sent to a bi-directional GRU (Bi-GRU).
The feature sequences are processed in both forward and reverse order and the results are concatenated at each time step.
Suppose the input feature at time step $t$ is $\mathbf{a}_t$, the forward pass of Bi-GRU performs the following operations at time $t$, and vice versa for the reverse pass:
\begin{equation}
\begin{tabular}{ll}
    $z_t  =\operatorname{sigmoid}(\mathbf{W}_z \cdot[ \mathbf{h}_{t-1}, \mathbf{a}_t])~;$ &  $r_t  =\operatorname{sigmoid}(\mathbf{W}_r \cdot[ \mathbf{h}_{t-1}, \mathbf{a}_t])~;$ \\
    $\tilde{ \mathbf{h}}_t  =\tanh ( \mathbf{W} \cdot[r_t *  \mathbf{h}_{t-1},  \mathbf{a}_t])~;$ & $\mathbf{h}_t =(1-z_t) *  \mathbf{h}_{t-1}+z_t * \tilde{ \mathbf{h}}_t~;$
\end{tabular}
\end{equation}
where $\mathbf{W}_z$, $\mathbf{W}_r$, $\mathbf{W}$ are learnable weights and $\mathbf{h}_t$ is the hidden state.
The sequence-level feature $\mathbf{F}\in\mathbb{R}^{D}$ is finally fed into a Dropout layer followed by a linear classifier with a softmax activation to compute the output probabilities $\mathbf{P}\in\mathbb{R}^{B\times R}$.

\subsection{Training Objective}\label{sec:objective}
In this work, we treat the optimal resolution determination at each bitrate as an individual classification task.
Therefore, let $\theta$ denote all the trainable parameters and $\mathbf{p}_i$ denote the $i$-th row of $\mathbf{P}$, the overall objective is expressed as:
\begin{equation}
\theta^*=\argmin_\theta\sum_{i=1}^{B}\mathcal{L}\left(\mathbf{p}_i,\mathbf{y}_i^*\right)~.
\end{equation}
During ground-truth construction, we notice a class imbalance problem between different resolutions.
For example, rare videos achieve the best quality at lower resolutions (\textit{e.g.} 216p), especially at high bitrates.
Therefore, we adopt the focal loss~\cite{lin2017focal} as the task-level loss function $\mathcal{L}$, which has been proven effective in solving imbalanced classification by focusing on hard misclassified instances.

\section{Experiments}

\subsection{Dataset}
We use the Inter-4K~\cite{stergiou2022adapool} dataset for model training.
Inter-4K is a standard benchmark for video super-resolution and frame interpolation, containing a total of 1000 UHD/4K videos with a predefined train-validation-test split.
The videos are captured at 60 frames per second (fps) with a duration of 5 seconds.
We also evaluate our model on two standard 4K benchmarks SJTU~\cite{song2013sjtu} and UVG~\cite{mercat2020uvg}.
The SJTU benchmark contains 15 sequences captured at 30 fps and most of them last for 10s.
The frame rates of the 16 UVG videos are 50/120 fps while the durations vary from 2.5s to 12s.
Following \cite{paul2022efficient,katsenou2021efficient,zabrovskiy2018multi}, all the sequences are standardized by downscaling to 1080p and cached by lossless encoding at 24 fps with 4:2:0 chroma subsampling.

\subsection{Implementation Details}
We utilize the Lanczos filter with $\alpha=3$ for all the spatial downscaling and upscaling operations.
The ResNet-18~\cite{he2016deep} pre-trained on ImageNet~\cite{deng2009imagenet} is adopted for feature extraction with the last fully-connected layer removed, leading to $D=512$.
For each sequence, $T=10$ frames are uniformly sampled to feed into the network.
The multi-head attention contains $n=4$ groups of self-attention.
The Bi-GRU consists of 2 layers with 256 hidden units, and the dropout probability is 0.25.
The model is trained for 100 epochs with cosine annealing using the SGD optimizer with an initial learning rate of 0.01, momentum of 0.9, weight decay of 0.0005, and batch size of 8.
Our experiments are implemented with PyTorch on a single NVIDIA Tesla A100.

\subsection{Quantitative Results and Ablation Studies}
The quantitative evaluation regarding optimal resolution prediction and convex hull approximation are presented in Tab.~\ref{tab:ablation}.
In addition to accuracy, we also report the F-Score and the G-Mean, which are more suitable for imbalanced classification~\cite{gu2009evaluation}, to evaluate the prediction performance.
In brief, F-Score is the harmonic mean of precision and recall while G-Mean is the geometric mean of the class-wise sensitivity.
Besides, we adopt the BD-Rate and the BD-VMAF to reflect the average bitrate and quality differences between the estimated bitrate ladder with the optimal convex hull.
Tab.~\ref{tab:ablation} also provides an ablation study of the proposed method.
When the focal loss is replaced with the standard cross-entropy loss, although the accuracy improves by 0.7\%, the F-Score and G-Mean decrease, and the overall RD performance deteriorates, which justifies the benefits of focal loss on the less frequent optimal resolutions.
\begin{table}[tp]
\centering
\footnotesize
\begin{tabular}{ccc|ccc|cc}\toprule\footnotesize
TA & FL & IR & Accuracy {($\uparrow$)}&F-Score {($\uparrow$)}&G-Mean {($\uparrow$)}&BD-Rate {($\downarrow$)}&BD-VMAF {($\uparrow$)}\\\midrule
\checkmark &  &  &\textbf{0.8620}&0.8523&0.8410&1.585\%&-0.3024\\
& \checkmark &  &0.8190 & 0.7990 & 0.7532 &2.197\%& -0.5086\\
&\checkmark & \checkmark&0.8160 &0.8132&0.8027& 2.166\% &-0.4318\\
\checkmark & \checkmark & \checkmark&0.8310&0.8245&0.8101&1.848\%&-0.3809\\\midrule
\checkmark & \checkmark & & 0.8550&\textbf{0.8549}&\textbf{0.8564}&\textbf{1.206\%}&\textbf{-0.2236}\\\bottomrule
\end{tabular}
\vspace{-5pt}
\caption{\label{tab:ablation}%
Ablation study on the Inter-4K~\cite{stergiou2022adapool} test set. ``TA'': ``Temporal Attention'', ``FL'': ``Focal Loss'', ``IR'': ``Input Resize''. $\uparrow$ means the higher the better, and vice versa.}
\vspace{-5pt}
\end{table}

\begin{table}[tp]
    \centering\footnotesize
    \begin{tabular}{cc}
    \footnotesize
    \begin{tabular}{l|rr}\toprule
         \textbf{Sequence Title} &  Zabrovskiy~\cite{zabrovskiy2018multi} &  Ours\\\midrule
        Bund Nightscape &-16.81&\textbf{0.00}\\
        Campfire Party &-3.32&\textbf{-1.34}\\
        Construction Field &-15.22&\textbf{0.00}\\
        Fountains &-5.01&\textbf{-0.38}\\
        Library &-16.43&\textbf{0.00}\\
        Marathon &-2.89&\textbf{-1.08}\\
        Residential Building &-18.76&\textbf{0.00}\\
        Runners & -4.99&\textbf{-1.40}\\
        Rush Hour &-4.40&\textbf{-0.03}\\
        Scarf &-16.53&\textbf{-1.16}\\
        Tall Buildings &-23.84&\textbf{-0.36}\\
        Traffic Flow &-13.43&\textbf{-0.02}\\
        Traffic and Building &-16.28&\textbf{0.00}\\
        Tree Shade &-15.69&\textbf{0.00}\\
        Wood & -13.48&\textbf{-0.47}\\
        --&--&--\\\midrule
         \textbf{Average}&-12.47&\textbf{-0.42}\\\bottomrule
    \end{tabular}&
    \footnotesize%
    \begin{tabular}{l|rr}\toprule
        \textbf{Sequence Title} &  Zabrovskiy~\cite{zabrovskiy2018multi} &  Ours\\\midrule
        Beauty &-1.07&\textbf{-0.73}\\
        Bosphorus &-4.60&\textbf{-0.47}\\
        CityAlley &-19.47&\textbf{0.00}\\
        FlowerFocus &-5.42&\textbf{-0.35}\\
        FlowerKids &-8.14&\textbf{-0.46}\\
        FlowerPan &-14.51&\textbf{-0.16}\\
        HoneyBee &-8.25&\textbf{-0.08}\\
        Jockey &\textbf{-1.00}&-4.52\\
        Lips &-4.79&\textbf{-0.32}\\
        RaceNight &-4.07&\textbf{-1.34}\\
        ReadySteadyGo &-3.34&\textbf{-3.17}\\
        RiverBank &-12.53&\textbf{-0.05}\\
        ShakeNDry &-2.78&\textbf{-1.22}\\
        SunBath &-1.61&\textbf{-0.48}\\
        Twilight &-18.60&\textbf{-0.61}\\
        YachtRide &-3.46&\textbf{-1.19}\\\midrule
         \textbf{Average}&-7.10&\textbf{-0.95}\\\bottomrule
    \end{tabular}\\
    \vspace{-10pt}\\
    (a) The SJTU~\cite{song2013sjtu} sequences. & (b) The UVG~\cite{mercat2020uvg} sequences.\\
    \end{tabular}
    \vspace{-20pt}
    \caption{BD-VMAF performances on public test sequences, the higher the better.}
    \label{tab:comparisons}
    \vspace{-25pt}
\end{table}
From Tab.~\ref{tab:ablation}, we can also observe that the temporal attention module consistently facilitates the prediction performance.
Besides, since the 2D CNN is pre-trained at the input size of 224, we attempt to resize the frames from 1080p to 224p before feature extraction.
Experimental results show that such input resizing leads to inferior results, which indicates that keeping the original resolution would be helpful for understanding the spatial complexity for optimal bitrate ladder construction.

The RD performances on the benchmark sequences are reported in Tab.~\ref{tab:comparisons}.
We compare the BD-VMAF differences against the optimal convex hulls with the DASH ladder~\cite{zabrovskiy2018multi} that is also free of pre-encoding.
Across all the sequences, our method substantially outperforms the DASH ladder except for Jockey.
This sequence presents horse racing with fast camera motion, which suggests that there is still a potential to improve the proposed method for scenes with high temporal complexity.

Since the input data is fixed as a 10-frame chunk, the inference speeds are stable for all videos.
More specifically, the inference time of the proposed TAGRN is 0.067 seconds/video on GPU and 2.83 seconds/video on CPU, which is trivial compared to the saved pre-encoding times.
Moreover, the inference speed can be further improved, \textit{e.g.}, by using a more lightweight 2D CNN for spatial feature extraction.

\subsection{Bitrate Ladder Visualizations}

\begin{figure}[t]
    \centering
    \includegraphics[width=\linewidth]{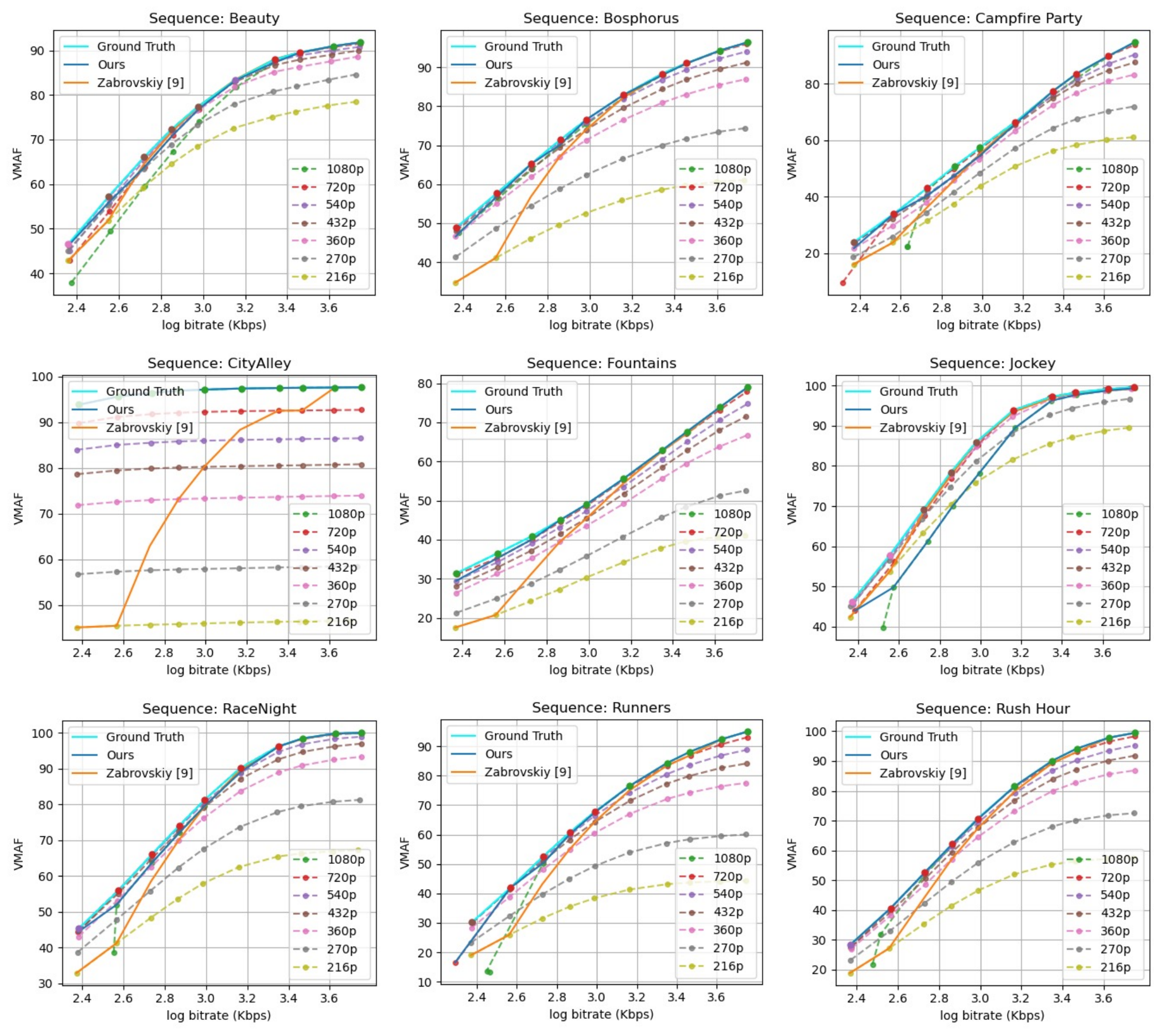}
    \vspace{-15pt}
    \caption{Plots of the RD curves of the predicted bitrate ladders on representative sequences.}
    \vspace{-5pt}
    \label{fig:rdcurve}
\end{figure}

The estimated bitrate ladders on representative sequences are plotted in Fig.~\ref{fig:rdcurve}, along with the ground-truth convex hulls and the RD curves obtained at each individual resolution.
Our results better approximate the optimal convex hulls compared with the DASH ladder~\cite{zabrovskiy2018multi} in most plots.

In theory, the overall distortion of downscaled encoding can be approximated with the summation of the rescaling error and the coding error~\cite{dong2013adaptive,wang2014adaptive}.
The practical RD performance on a video depends on its spatiotemporal characteristic and the resulting tradeoff between the rescaling and the coding errors.
Interestingly, on the sequences with dense edges and low temporal variations like CityAlley, the coding error is overwhelmed by the rescaling loss.
Therefore, encoding at higher resolution is always superior to lower resolution over the whole bitrate range for such videos.
In this case, content-agnostic streaming with low resolution at low bitrates as in the DASH ladder~\cite{zabrovskiy2018multi} would result in sub-optimal quality of experiences.
Worse still, the convex hull estimation methods based on intersection points between different resolutions~\cite{katsenou2021efficient,katsenou2021vmaf} turn out to be completely inapplicable.
This kind of method also fails on sequences with multiple intersection points between different resolutions such as Campfire Party, where some resolutions (\textit{e.g.} 1080p and 720p) alternately become the optimal as the bitrate increases.
Differently, our method still achieves satisfactory approximation to the Pareto fronts of such sequences.

\section{Conclusion}
In this paper, we introduce a novel framework for efficient content-aware bitrate ladder estimation by formulating it as a multi-task classification problem.
Different from existing approaches, our method is completely free of pre-encodings to determine the optimal transcoding resolutions at deployment, dramatically reducing the encoding costs.
We develop the TAGRN to capture spatial features from sampled frames that are subsequently aggregated and classified in the time domain. 
When compared to brute-force searching, our method shows an average BD-Rate loss of only 1.206\% and a BD-VMAF loss of -0.2236 but achieves the total elimination of pre-encodings.

\Section{References}
\bibliographystyle{IEEEbib}
\bibliography{refs}

\end{document}